\documentclass[conference]{IEEEtran}
\IEEEoverridecommandlockouts
\usepackage{amsthm}
\usepackage{amsfonts,amsmath}
\usepackage{algorithm}
\usepackage[noend]{algpseudocode}
\usepackage{bm}
\usepackage{booktabs}
\usepackage{soul}
\usepackage{multirow}
\usepackage{subfig}
\usepackage[flushleft]{threeparttable}
\usepackage{adjustbox}
\usepackage{fancyhdr}
\usepackage[dvipsnames]{xcolor}
\usepackage[flushleft]{threeparttable}
\usepackage{tikz}
\usepackage{outlines}
\usepackage{balance}
\usepackage{tcolorbox}
\usepackage{wrapfig}
\usepackage{blindtext}
\usepackage{xcolor}
\usepackage{balance}
\usepackage{algorithm}
\usepackage{color}
\usepackage{listings}
\usepackage{color}
\usepackage{MnSymbol,wasysym}
\usepackage{marvosym}
\usepackage[switch]{lineno}
\usepackage{array}
\usepackage[switch]{lineno}
\usepackage{tabularx}
\usepackage{array} 
\newcolumntype{Y}{>{\centering\arraybackslash}X} 
\newcolumntype{Z}{>{\centering\arraybackslash}p{2cm}} 

\usepackage{cite}
\usepackage{MnSymbol,wasysym}
\usepackage{marvosym}
\usepackage{times}
\usepackage{caption}
\usepackage{textcase}

\usepackage{eso-pic}

\usepackage{xcolor,colortbl}
\def\BibTeX{{\rm B\kern-.05em{\sc i\kern-.025em b}\kern-.08em
    T\kern-.1667em\lower.7ex\hbox{E}\kern-.125emX}}

\begin{document}
\bstctlcite{IEEEexample:BSTcontrol}
\title{Exploiting On-chip Heterogeneity of Versal Architecture for GNN Inference Acceleration

}


\author{
\IEEEauthorblockN{Paul Chen\textsection, Pavan Manjunath\textsection, Sasindu Wijeratne,  Bingyi Zhang, Viktor Prasanna}
\IEEEauthorblockA{
    University of Southern California, Los Angeles, CA, USA \\
    \{chenpaul, pavanman,  kangaram, bingyizh, prasanna\}@usc.edu}
}


\maketitle

\begingroup\renewcommand\textsuperscript{\textsection}
\footnotetext{\textsection Equal contribution}
\endgroup

\vspace{-5mm}
\begin{abstract}

Graph Neural Networks (GNNs) have revolutionized many Machine Learning (ML) applications, such as social network analysis, bioinformatics, etc. GNN inference can be accelerated by exploiting data sparsity in the input graph, vertex features, and intermediate data in GNN computations. 
For dynamic sparsity exploitation, we leverage the heterogeneous computing capabilities of AMD Versal ACAP architecture to accelerate GNN inference. We develop a custom hardware module that executes the sparse primitives of the computation kernel on the Programmable Logic (PL) and efficiently computes the dense primitives using the AI Engine (AIE). To exploit data sparsity during inference, we devise a runtime kernel mapping strategy that dynamically assigns computation tasks to the PL and AIE based on data sparsity. Our implementation on the VCK5000 ACAP platform leads to superior performance compared with the state-of-the-art implementations on CPU, GPU, ACAP, and other custom GNN accelerators. Compared with these implementations, we achieve significant average runtime speedup across various models and datasets of 162.42×, 17.01×, 9.90×, and 27.23×, respectively. Furthermore, for Graph Convolutional Network (GCN) inference, our approach leads to a speedup of 3.9-96.7× compared to designs using PL only on the same ACAP device.

\end{abstract}


\vspace{-3mm}
\section{Introduction}

\vspace{-1mm}
Graph Neural Networks (GNNs) have become increasingly popular in recent years due to their ability to effectively learn from (unstructured) graph data. GNNs offer remarkable versatility and can be applied to a wide range of graph-related problems, including node classification \cite{nodec}, link prediction \cite{NEURIPS2018_53f0d7c5}, graph classification \cite{zhang2022accurate}, etc. This versatility has established GNNs as a powerful technique for various domains, such as computer vision \cite{pradhyumna2021graph}, natural language processing \cite{wu2021deep}, and recommendation systems \cite{gao2022graph}, among others. In many practical applications \cite{dou2020enhancing}, performing low-latency GNN inference is crucial for enabling real-time decision-making.


The computational characteristics of GNN inference present challenges for real-time applications, primarily due to the high computational complexity and the irregular memory access of graph data. CPUs are ill-suited for GNN acceleration \cite{dgl-cpu} due to their sequential instruction-based architecture. On the other hand, GPUs excel at parallel processing and can accelerate GNNs. Still, they have limitations (e.g., complex cache hierarchy) in handling certain graph structures and memory access requirements \cite{wang2021gnnadvisor}.
To address these challenges, Field-Programmable Gate Arrays (FPGAs) offer a compelling solution. FPGAs provide flexibility \cite{meng2021dynamap, meng2022accelerator}, programmability, and parallelism \cite{meng2023framework, ye2020hybriddnn, zhang2020dnnexplorer}, making them well-suited for specific tasks such as message passing in GNN inference. 


The Adaptive Compute Acceleration Platform (ACAP)\cite{acap} offers sequential instruction-based execution, parallel vector processing, and adaptive computing. Because GNN computation kernels can be mapped to sparse and dense primitives based on dynamic sparsity exploitation \cite{zhang2023dynasparse}, ACAP offers a promising platform for accelerating GNN inference. The programmable logic (PL) component of ACAP can be leveraged to handle sparse primitives, while the AI Engine (AIE) are well suited to handle dense primitive. Nonetheless, there are several challenges in achieving efficient GNN acceleration using ACAP: (1) Developing an efficient hardware module for PL is crucial to accelerate  sparse primitives effectively---this module must be carefully designed and optimized to maximize performance and resource utilization.
(2) Although AI Engine exhibits high peak performance, achieving low-latency inference, using them can be a complex undertaking. Optimizing the utilization of the AI Engine and minimizing inference latency requires careful consideration of algorithmic optimizations to exploit the architectural features.
(3) The interaction between PL and AIE must be  designed efficiently to reduce data communication overhead. Effective data movement and synchronization mechanisms need to be implemented to facilitate seamless collaboration between the PL and AIE.
The key contributions of this paper are:
\begin{itemize}
\item We develop an efficient accelerator design that leverages the heterogeneity of PL and AIE  of the Versal architecture to accelerate GNN inference. The accelerator executes sparse primitives on PL and dense primitive on the AIE.

\item We develop a runtime system that consists of a task analyzer and scheduler using the on-chip ARM processor that dynamically assigns computation tasks to the PL and AIE based on data sparsity.

\item We evaluate the design on diverse datasets, including CiteSeer (CI), Cora (CO), PubMed (PU), Flickr (FL), NELL (NE), and Reddit (RE), for inference using  state-of-the-art GNN models such as GCN, GraphSage, GIN, and SGC. The experimental results show that our implementation on VCK5000 achieves 162.42×, 17.01×, 9.90×, and 27.23× average speedup compared with the state-of-the-art CPU, GPU, ACAP, and other custom GNN accelerators, respectively.
\end{itemize}

The rest of the paper is organized as follows: Section \ref{sec:background} introduces the Background and Related work. In Section \ref{sec:accelerator-design}, we demonstrate the intricate details of the Accelerator's design. The evaluation results are presented in Section \ref{sec:experimental-results}. Finally, we conclude the paper in Section V.



\section{Background and Related Work}
\label{sec:background}

\subsection{Background}

\subsubsection{Graph Neural Networks}

GNNs have been proposed for representation learning on graphs denoted as $\mathcal{G}(\mathcal{V}, \mathcal{E})$. GNNs follow the message-passing paradigm (as outlined in Algorithm \ref{alg:GNN-computation-abstraction}), where vertices recursively aggregate information from their neighbors. The last-layer embedding of the target vertex $v$ is denoted as $\bm{h}^{L}_{v}$. Typically, the Update() operation is implemented as a Multi-Layer Perceptron that transforms the vertex features.
After the Aggregate() and Update() operations in each layer, an element-wise activation function is applied to the feature vectors. The output embedding $\bm{h}^{L}_{v}$ can be utilized for various downstream tasks, including node classification (\cite{hamilton2017inductive,kipf2016semi}), link prediction, and more.
GCN \cite{kipf2016semi}, GraphSAGE \cite{hamilton2017inductive}, GIN \cite{xu2018powerful}, and SGC \cite{wu2019simplifying} are some representative GNN models. 
 
\vspace{-1mm}
\begin{algorithm}
\caption{GNN Computation Abstraction}
\label{alg:GNN-computation-abstraction}
\begin{small}
\begin{algorithmic}[1]
 \renewcommand{\algorithmicrequire}{\textbf{Input:}}
\renewcommand{\algorithmicensure}{\textbf{Output:}}
 \Require Input graph: $\mathcal{G}(\mathcal{V},\mathcal{E})$; vertex features: $\left\{\bm{h}^{0}_{1}, \bm{h}^{0}_{2}, \bm{h}^{0}_{3}, ..., \bm{h}^{0}_{|\mathcal{V}|}\right\}$;
 \Ensure Output vertex features $\left\{\bm{h}^{L}_{1}, \bm{h}^{L}_{2}, \bm{h}^{L}_{3}, ..., \bm{h}^{L}_{|\mathcal{V}|}\right\}$;
\For{$l=1...L$}
\For{each vertex $v \in \mathcal{V}$}
\State{$\bm{a}_{v}^l = {\text{Update}(}\bm{h}_{v}^{l-1}, \bm{W}^{l} \textbf{)}$}
\State{$\bm{z}^l_{v} = {\text{Aggregate}(}\bm{a}_{u}^{l}: u\in \mathcal{N}(v))$}
\State{$ \bm{h}_{v}^l = \sigma(\bm{z}_{v}^l )$}
\EndFor
\EndFor
\end{algorithmic}
\end{small}
\end{algorithm}

\vspace{-3mm}
\subsubsection{Computation Kernels and Primitives in GNNs}\label{compute_kernels_}

The computation kernels involved in GNN inference consist of feature aggregation and feature transformation, corresponding to the Aggregate() and Update() operations in the message-passing paradigm of GNNs. These computation kernels can be mapped to fundamental computation primitives based on the data sparsity. These primitives include dense-dense matrix multiplication (GEMM), sparse-dense matrix multiplication (SpDMM), and sparse-sparse matrix multiplication (SpMM). 
\vspace{-0.5mm}
\subsection{Data Sparsity in GNN Inference}
\label{subsec:GNN-sparsity}

The \emph{density} of a matrix is defined as the total number of non-zero elements divided by the total number of elements. Note that, the \emph{sparsity} is given by $(1 - \text{\emph{density}})$. The computation kernels in GNNs involve three types of matrices: graph adjacency matrix $\bm{A}$, vertex feature matrix $\bm{H}$, and weight matrix $\bm{W}$. 
The adjacency matrix $\bm{A}$ of different graph datasets \cite{pyg-dataset} can have different densities. For a given adjacency matrix, different parts of the matrix can have different densities. 
For various graphs, the input feature matrices can have different densities. The feature matrices of different layers also have different densities.  For the weight matrices, prior works \cite{rahman2022triple, chen2021unified} have  proposed various pruning techniques to reduce the density of the weight matrices. 
To leverage the above data sparsity, Zhang et al. \cite{zhang2023dynasparse} propose a technique called Dynasparse, which focuses on dynamically mapping computation kernels to primitives such as GEMM, SpDMM, and SpMM. The authors introduce a unified hardware architecture capable of supporting various primitives (GEMM, SpDMM, SpMM). This architecture offers different execution modes, each with distinct computation parallelism and the ability to skip zero-elements in the input matrix.
Furthermore, the authors develop a runtime system that dynamically maps computation kernels to the appropriate primitives (to be executed on the unified architecture) using a performance model based on data sparsity. The performance model considers the trade-off between the computation parallelism and the ability to skip zero-elements of different execution modes, in order to reduce the inference latency. In this study, we extend the dynamic kernel-to-primitive mapping strategy from Dynasparse \cite{zhang2023dynasparse}  to leverage the heterogeneous computing capabilities of the ACAP architecture for accelerating GNN inference. Specifically, for hardware mapping, we employ the AIE to execute the dense primitive (GEMM) due to its high peak performance. Additionally, we utilize the PL to construct a customized data path and memory organization, enabling efficient execution of sparse primitives (SpDMM, SpMM).

\vspace{-3mm}
\subsection{Related Work}

H-GCN\cite{zhang2022h} introduces a hybrid accelerator that leverages the heterogeneity of ACAP architecture by partitioning the input graph into subgraphs and assigning computations to either the AI Engine (AIE) or the Programmable Logic (PL) based on subgraph density. However, H-GCN's graph partitioning and reordering approach can result in significant preprocessing overhead.
In contrast, our work adopts a simple data partitioning strategy where we decompose the GNN kernel into different primitives (Section~\ref{compute_kernels_}) and dynamically map them to the AIE or PL based on the data sparsity at runtime. This approach eliminates the need for complex graph partitioning and enables efficient execution of the GNN computations.
The Dynasparse framework~\cite{zhang2023dynasparse} presents a hardware-software codesign for accelerating GNN inference on data-center FPGAs. It encompasses offline compilation optimizations, a runtime system based on soft processors, and a PL-based accelerator design that exploits sparsity. In contrast, our work capitalizes on the heterogeneity of AMD ACAP devices, utilizing an ARM Cortex-A72 processor to execute a runtime system. Additionally, we employ both the PL and AIE components to execute GNN kernels mapping to different primitives of the GNN inference computations, leveraging the specific strengths of each component.
Several existing works~\cite{geng2021gcn, geng2020awb, yan2020hygcn, zhang2021boostgcn, liang2020deepburning, sarkar2022flowgnn, zhang2020hardware, zhang2023graphagile, zhang2022low, zhang2023dynasparse, lin2022hp, zhang2021efficient, lin2021gcn, zhou2022model, zhang2020accelerating} have proposed FPGA-based acceleration techniques for GNN inference without AIE. These works typically employ custom compute hardware modules for operations such as SpMM, SpDMM, and GEMM on PL. In contrast, our work focuses on mapping onto the most suitable hardware components in ACAP to execute these compute kernels efficiently. By leveraging the capabilities of both PL and AIE, we enable efficient GNN inference on the ACAP platform.

 

\vspace{-3mm}
\section{Accelerator Design}
\label{sec:accelerator-design}


\subsection{Problem Definition}

Our objective is to leverage the computational characteristics of the Programmable Logic - AI Engine, along with the Processor System (PS) of ACAP architecture, to accelerate full graph inference. Full graph inference involves performing the message-passing paradigm (as described in Algorithm \ref{alg:GNN-computation-abstraction}) on the entire graph \cite{geng2020awb, yan2020hygcn, zhang2021boostgcn, zhang2023dynasparse}. This can be computationally demanding and memory-intensive, particularly for large graphs which do not fit on the on-chip memory.

To address this challenge, we propose an accelerator that effectively utilizes the on-chip heterogeneity of ACAP platform. By leveraging both the PL and AI Engine, our accelerator can efficiently accelerate GNN inference on datasets with varying degrees of sparsity. Note that our approach does not require generating an accelerator for each input graph and GNN model, thereby enhancing its efficiency and flexibility. For a given input graph and GNN model, initially stored on the host memory, we perform pre-processing (Section \ref{preprocessing}) of the input graph and the GNN model on the host processor for hardware execution  and transfer the processed input graph and GNN model to FPGA DDR.

\vspace{-1mm}
\subsection{System Overview}
\label{subsec:system-overview}
\vspace{-5mm}
\begin{figure}[h]
  \centering
  \includegraphics[width=0.9\linewidth]{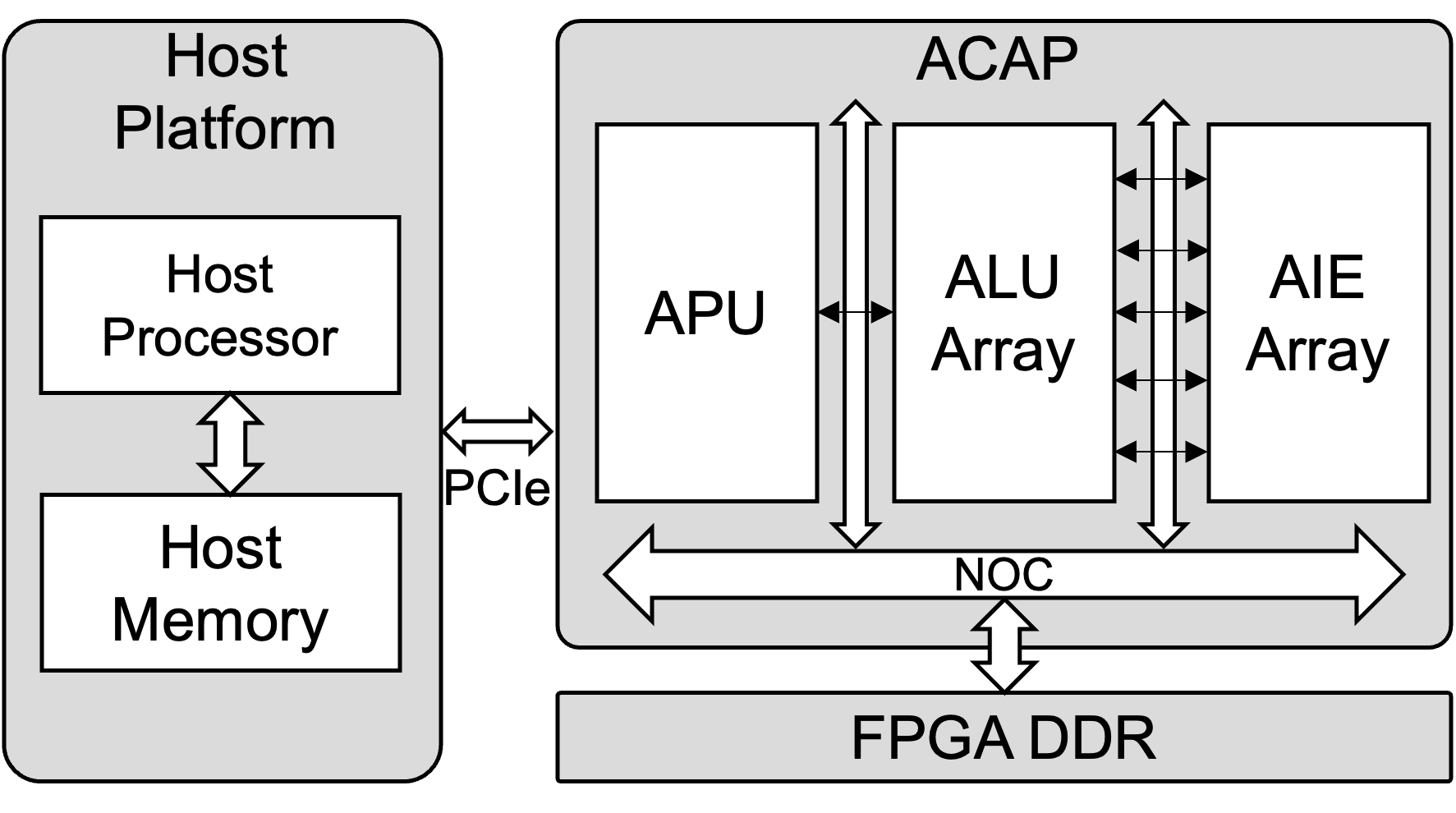}
  \caption{System Overview}
  \label{fig:overview}
\end{figure}
\vspace{-2mm}



Figure \ref{fig:overview} depicts the proposed design of leveraging ACAP architecture for dynamic sparsity exploitation (See Dynasparse \cite{zhang2023dynasparse}) in GNN inference.
The architecture consists of three main parts: Application Processing Unit (APU), Programmable Logic (PL), and AI Engine (AIE) array. On PL, we implement multiple ALU (Arithmetic Logic Unit) arrays to execute sparse primitives (SpDMM, SpMM). The AIE array efficiently executes the dense primitive (GEMM) due to low-latency inter-tile communication and high computation density. The APU hosts a runtime system that dynamically maps kernels for execution. The host processor performs preprocessing for the input graph and GNN model. The board also has a high-performance communication infrastructure that efficiently interconnects computational and memory elements called Network on Chip (NoC). The input graph and the GNN model are stored in the host memory. After preprocessing, they are transferred to the FPGA DDR.

\vspace{0.1cm}
\label{preprocessing}
\noindent \textbf{Preprocessing}: For preprocessing, the host processor performs 2-D data partitioning \cite{10.1007/978-3-642-14122-5_15}, partitioning the input graphs into smaller submatrices along both dimensions to fit in the on-chip memory of ACAP, and enable parallel processing and efficient computation, for feature matrix $\bm{H}$, graph adjacency matrix $\bm{A}$, and weight matrix $\bm{W}$. We use $\bm{X}_{ij}$ to denote a partition of matrix $\bm{X}$. 

\vspace{0.1cm}
\noindent \textbf{Runtime}: The runtime system consists of an \emph{analyzer} and a \emph{scheduler}. The analyzer dynamically maps the computation kernels (e.g., feature aggregation, feature transformation) to the basic primitives (GEMM, SpDMM, and SpMM) based on the data sparsity. As the AIE array is efficient for dense primitives and ALU arrays are efficient for sparse primitives, the analyzer uses a performance model to determine the kernel-to-primitive mapping and creates the tasks. Then, the scheduler adds the tasks to the task queues and dynamically schedules the tasks to ALU arrays and AIE array.

\begin{figure*}[!t]
  \centering
  \includegraphics[width=1\textwidth]{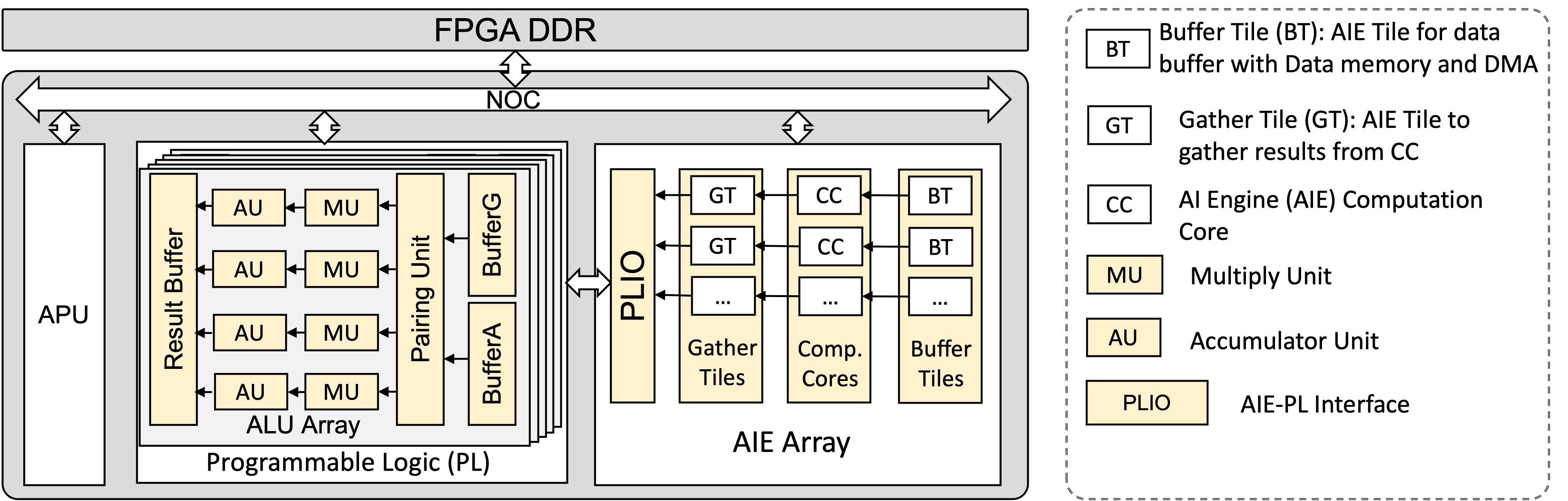}
  \caption{Overall Accelerator Design}
  \label{fig:arch}
\end{figure*}

The following two sections elaborate on the hardware design and the runtime system. Figure \ref{fig:arch} depicts the details of the proposed accelerator.

\vspace{-1mm}
\subsection{AI Engine (AIE) Array}




The AI Engine Array is responsible for executing the dense-dense matrix multiplication (GEMM). Figure \ref{fig:arch} provides an illustration of the organization of the AIE array specifically designed for GEMM execution. It consists of three main components: Buffer Tiles (BTs), Computation Cores (CCs), and Gather Tiles (GTs).
To execute a GEMM operation $\bm{X} \times \bm{Y}$, the BTs load the input matrices, denoted as $\bm{X}$ and $\bm{Y}$, into their data memory from the DDR through the Direct Memory Access (DMA) engine. The loaded data is then transferred to the CCs.
Communication between two consecutive kernels is established using a common buffer in the shared memory module \cite{xilin_ai_engine}. Neighboring AI engine tiles can easily share data without memory transfers over DMA and AXI4-Stream interconnect by using the shared memory.

\begin{figure}[h]
  \centering
  \includegraphics[width=0.9\linewidth]{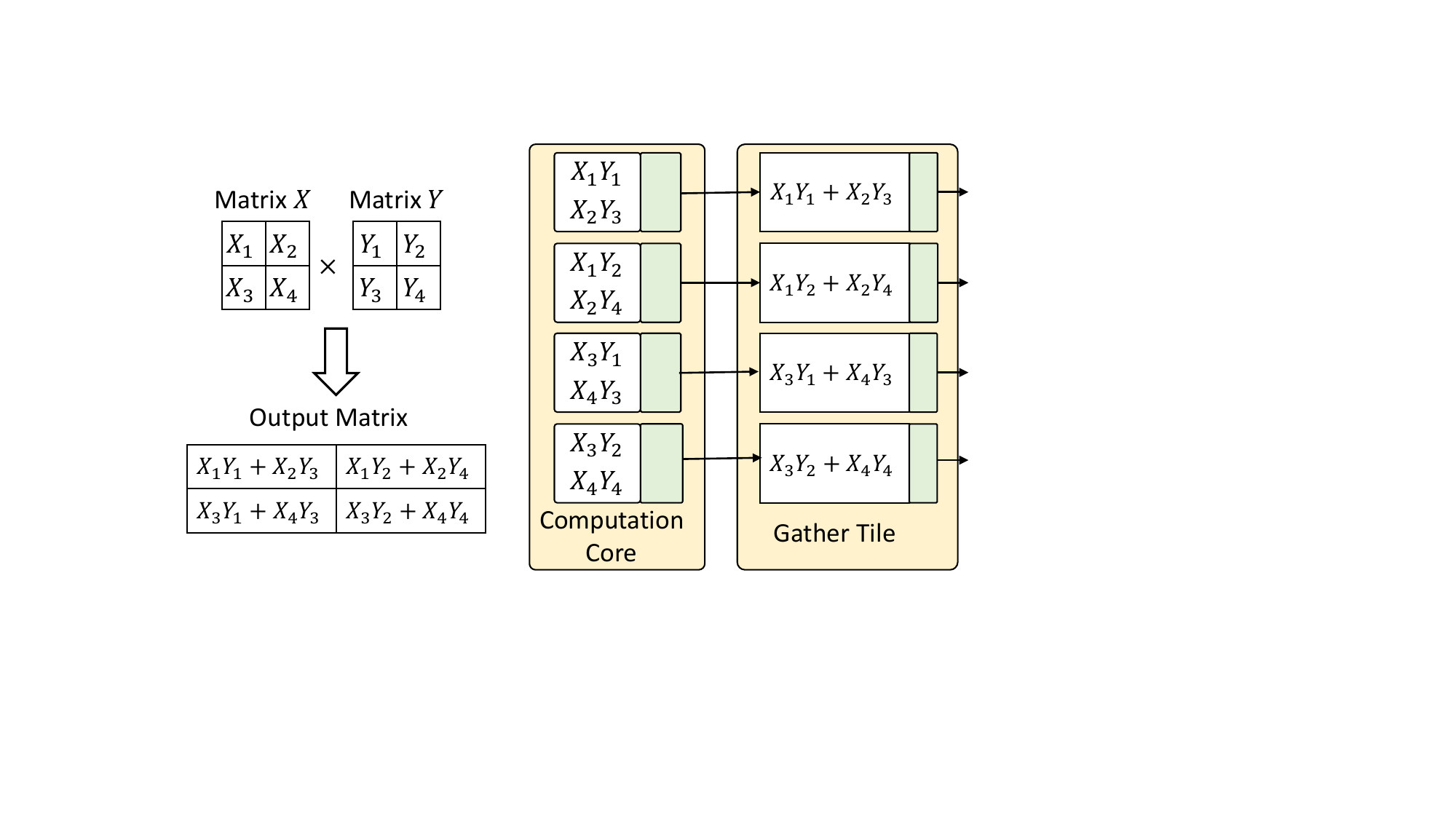}
  \caption{AI Engine Computation Core}
  \label{fig:AIE-CC}
  \vspace{-5mm}
\end{figure}



\vspace{0.1cm}
\noindent \textbf{AIE Computation Core (CC)}: 
Each AIE Computation Core (CC) consists of four AIE tiles. Each AIE tile is equipped with its own data memory module. The data flow involves transferring the data from the Buffer Tiles to the AIE Tiles.
During each cycle, the vertex feature vectors are loaded into the AIE tile. The next step involves performing the Multiply-Accumulation (MAC) operation using the partial results obtained in the previous cycle. Figure \ref{fig:AIE-CC} illustrates the computation process of executing the matrix multiplication $\bm{X} \times \bm{Y}$ on a Computation Core.
Each matrix (e.g., $\bm{X}$, $\bm{Y}$) is evenly divided into four partitions. In each cycle, the $\bm{X}$ matrix is loaded in row-major order, and the $\bm{Y}$ matrix is loaded in column-major order into the respective CC. In the first cycle, the first row of matrix data is multiplied, and in the consequent cycles, subsequent rows are multiplied and accumulated with the previous product, as shown in the output matrix in Figure \ref{fig:AIE-CC}. The final output is sent to the Gather AIE Tiles to form the output matrix.

\vspace{-0.5mm}
\subsection{Arithmetic Logic Unit (ALU) Array}

The sparse primitives, specifically SpDMM and SpMM, are executed on the ALU Arrays, which are designed for efficient execution of sparse matrix multiplication.
Figure \ref{fig:arch} illustrates the architecture of the Arithmetic Logic Unit (ALU) Array. This array consists of $p$ computation pipelines, each comprising a Multiply Unit (MU) and an Accumulator Unit (AU). Each Multiply Unit contains an array of $q$ hardware multipliers, while each Accumulator Unit consists of an array of $q$ accumulators. Each Multiply unit and accumulator is instantiated as a Digital Signal Processing (DSP) slice on FPGA and value of $p$ and $q$  is restricted by the number of DSPs available. Additionally, the ALU Array incorporates three data buffers: BufferA, BufferG, and Result Buffer (RB). BufferA and BufferG are responsible for storing the two input matrices, denoted as $\bm{X}$ and $\bm{Y}$ respectively, while the Result Buffer stores the output matrix, $\bm{Z}$. To facilitate the routing of input data from BufferA and BufferG to the computation pipelines, each ALU Array includes a Pairing Unit. The Pairing Unit for each non-zero element in $\bm{X}$ from BufferA it fetches $q$ elements from BufferG. It effectively handles the irregular memory access patterns typically associated with sparse primitives. Furthermore, the ALU Array operates in two distinct execution modes: SpDMM mode and SpMM mode, dedicated to the execution of SpDMM and SpMM, respectively. The execution mode is set by the control
bits of the ALU array. The overhead of switching execution modes is just one clock cycle.

\begin{algorithm}
\caption{SpDMM using Scatter-Gather Paradigm}\label{alg:scatter-gather}
\begin{small}
\begin{algorithmic}[1]
\renewcommand{\algorithmicrequire}{\textbf{Input:}}
\renewcommand{\algorithmicensure}{\textbf{Output:}}
 \Require Sparse matrix (BufferA): $\bm{X}$; Dense matrix (BufferG): $\bm{Y}$;
 \Ensure Output matrix  (Result Buffer): $\bm{Z}$ ($\bm{Z} = \bm{X}\times \bm{Y}$);
\While {not done}
\For{each $e(i,j,value)$ in $\bm{X}$ \textbf{Parallel}}  {\color{blue}\Comment{Scatter Phase}}
\State Fetch $\bm{Y}[i]$ from BufferG   {\color{blue}\Comment{Pairing Unit}}
\State Form input pair ($\bm{Y}[i]$, $e$)  {\color{blue}\Comment{Pairing Unit}}
\EndFor

\For{each input pair \textbf{Parallel}}
{\color{blue}\Comment{Gather Phase}}
\State $u \gets ${Update($\bm{Y}[i],e.value$)}  {\color{blue}\Comment{Multiply Unit}}
\State Fetch $\bm{Z}[j]$ from Result Buffer
\State {$\bm{Z}[j] \gets$ {Reduce($u$)}}   {\color{blue}\Comment{Accumulator Unit}}
\EndFor
\EndWhile
\end{algorithmic}
\end{small}
\end{algorithm}

\noindent \textbf{SpDMM Mode}: Multiplication of a sparse matrix with a dense matrix is executed using the Scatter-Gather Paradigm \cite{zhang2023dynasparse} shown in Algorithm \ref{alg:scatter-gather}. The sparse matrix denoted as $\bm{X}$ is stored in BufferU using the Coordinate (COO) format. The dense matrix  denoted as $\bm{Y}$ is stored in BufferG. In SpDMM Mode the ALU array can execute upto $p \times q$ MAC operations per clock cycle.

\begin{algorithm}
\caption{SpMM using Row-wise Product}\label{alg:scatter-gather-SPMM}
\begin{small}
\begin{algorithmic}[1]
\renewcommand{\algorithmicrequire}{\textbf{Input:}}
\renewcommand{\algorithmicensure}{\textbf{Output:}}
 \Require  Sparse matrix (BufferA): $\bm{X}$; Sparse matrix (BufferG): $\bm{Y}$;
 \Ensure Output matrix (In Result Buffer): $\bm{Z} = \bm{X}\times \bm{Y}$;
\For{each row $\bm{Z}[j]$ in $\bm{Z}$ \textbf{Parallel}}
\State Assign the workload of $\bm{Z}[j]$ to $(j\%p)^{\text{th}}$ pipeline
\For{each $e(i,j,value)$ in $\bm{X}[j]$ }  {\color{blue}\Comment{Scatter Phase}}
\State Fetch $\bm{Y}[i]$ from BufferO   {\color{blue}\Comment{Pairing Unit}}
\State Form input pair ($\bm{Y}[i]$, $e$)  {\color{blue}\Comment{Pairing Unit}}
\EndFor

\For{each input pair ($\bm{Y}[i]$, $e$)}
{\color{blue}\Comment{Gather Phase}}
    \For{each non-zero $\bm{Y}[i][k]$ in $\bm{Y}[i]$}  
    \State Produce $u \gets $ {Update($e.value \times \bm{Y}[i][k]$)}
    \State {Merge $\bm{Z}[j][k] \gets$ {Reduce($u$)}} 
    \EndFor
    \EndFor
\EndFor
\end{algorithmic}
\end{small}
\end{algorithm}

\noindent \textbf{SpMM Mode}: The multiplication of two input sparse matrices is executed using the Row-wise Product  with Scatter-Gather paradigm as shown in Algorithm \ref{alg:scatter-gather-SPMM}. 
For Row-wise Product,  a row $ \bm{Z}[j]$  of  output matrix $\bm{Z}$ is calculated through: 
\begin{equation}
    \bm{Z}[j] = \sum_{i}\bm{X}[j][i]*\bm{Y}[i]
    \label{eq:workload-of-an-row}
    \vspace{-3mm}
\end{equation}
For calculating the output matrix $\bm{Z}$, a pipeline is assigned the workload of a row of output matrix (Equation \ref{eq:workload-of-an-row}). $p$ pipelines can calculate $p$ output rows in parallel until all the rows of the output matrices are calculated. SpMM Mode can execute $p$ multiply-accumulate (MAC) operations per clock cycle.


\subsection{Dynamic Task Management (Runtime System)}
\label{task scheduling}
In the proposed accelerator design, the AIE array is efficient for dense primitives(GEMM), and the ALU array is efficient for sparse primitives(SpDMM, SpMM).
To exploit various data sparsity in GNN inference, we implement a runtime system on APU that performs dynamic task management based on data sparsity.

Given a matrix multiplication $\bm{Z} = \bm{X} \times \bm{Y}$, we define a \emph{task} as the process of calculating the partition of the large output matrix $\bm{Z}$. For example, for a partition $\{\bm{Z}_{ij}\}$, the task can be expressed as:
\begin{equation}
    \bm{Z}_{ij} = \sum_{k}{\bm{X}_{ik}\times \bm{Y}_{kj}} \label{eq:computation-task}
    \vspace{-2mm}
\end{equation}
Therefore, each computation kernel (e.g., feature aggregation or feature transformation) can be decomposed into independent tasks. To execute these tasks efficiently, we dynamically schedule the tasks by analyzing the sparsity in the runtime system as shown in Algorithm \ref{alg:Dynamic-Task-Scheduling}.

\begin{algorithm}
\caption{Runtime System}
\label{alg:Dynamic-Task-Scheduling}
\begin{small}
\begin{algorithmic}[1]
 \renewcommand{\algorithmicrequire}{\textbf{Input:}}
\renewcommand{\algorithmicensure}{\textbf{Output:}}
 \Require  Input graph $\mathcal{G}(\mathcal{V}, \mathcal{E}, \bm{X})$; GNN model with $L$ layers;
 \Ensure Output of GCN Inference;
\State{$STQ \leftarrow \emptyset$} {\color{blue}\Comment{Sparse task queue}}
\State{$DTQ \leftarrow \emptyset$} {\color{blue}\Comment{Dense task queue}}
 \State {\color{brown} ======= Analyzer ========} 
\For{$l = 1$ to $L$}  {\color{blue}\Comment{Iterate each layer}}
    \For{ each kernel $kernel_{j}$ in layer $l$}  {\color{blue}\Comment{Iterate each kernel}}
        \For{each task $task_{k}$ in $kernel_{j}$}
            \State{$t_{\text{ALU}} = \mathcal{P}_{\text{ALU}}(task_{k})$ }
            \State{$t_{\text{AIE}} = \mathcal{P}_{\text{AIE}}(task_{k})$ }
            \If{$t_{\text{ALU}} > t_{\text{AIE}}$} 
                \State{$STQ.push(task_{k})$}
            \Else
                \State{$DTQ.push(task_{k})$}
            \EndIf 
        \EndFor
    \EndFor
\EndFor
\State {\color{brown} ======= Scheduler ========} 
\While{True}
    \If{there is an idle ALU array: $ALU_{i}$}
        \State{$task_{k} \leftarrow STQ.pop()$}
        \State{Execute $task_{k}$ on $ALU_{i}$}
    \EndIf
\EndWhile

\While{True}
    \If{AIE array is idle}
        \State{$task_{k} \leftarrow DTQ.pop()$}
        \State{Execute $task_{k}$ on AIE array}
    \EndIf
\EndWhile
\end{algorithmic}
\end{small}
\end{algorithm}
The runtime system consists of an \emph{Analyzer} and a \emph{Scheduler}. Moreover, the runtime system maintains two task queues -- Sparse Task Queue (STQ) and Dense Task Queue (DTQ). For each task, the analyzer estimates its execution time on the ALU array and AIE array based on the theoretical performance model $\mathcal{P}_{\text{ALU}}()$ and $\mathcal{P}_{\text{AIE}}()$. Then, the analyzer adds the task to the Sparse Task Queue  (STQ) or Dense Task Queue (DTQ) according to the estimated execution time. The scheduler schedules the tasks from the two task queues to the ALU arrays and AIE array.

\vspace{0.1cm}
\noindent \textbf{Performance Model ($\mathcal{P}_{\text{ALU}}()$, $\mathcal{P}_{\text{AIE}}()$):} 
The performance model (shown in Table \ref{Performance Model}) estimates the execution cycle of each task. Based on the performance model, the Analyzer adds the task to the appropriate task queue. We define the density $\alpha_{\bm{X}}$ of a matrix $\bm{X}$ as the total number of non-zero elements divided by the total number of elements. For a given task (Equation \ref{eq:computation-task}), we use $\bm{X}_{i,:}$ to denote the concatenation of $\{\bm{X}_{ik}\}$ and use $\bm{Y}_{:,j}$ to denote the concatenation of $\{\bm{Y}_{kj}\}$. Then, the task (Equation \ref{eq:computation-task}) can be rewritten as:
\begin{equation}
    \bm{Z}_{ij} = \bm{X}_{i,:} \times \bm{Y}_{:,j}
    \vspace{-1mm}
\end{equation}
which has several properties w.r.t data sparsity, including the density of $\bm{X}_{i,:} $ denoted as $\alpha_{\bm{X}_{i,:}}$, and the density of $\bm{Y}_{:,j}$  denoted as $\alpha_{\bm{Y}_{:,j}}$. Suppose $\bm{X}_{i,:}\in \mathbb{R}^{m\times n}$ and $\bm{Y}_{:,j}\in \mathbb{R}^{n\times d}$. Using AIE array to execute this task, the task is treated as GEMM, and the total execution cycle is:
\begin{equation}
    t_{\text{AIE}} =  \frac{mnd}{f_{\text{AIE}} * N_{\text{AIE}} * \beta}
\end{equation}
where $f_{\text{AIE}}$ is the frequency of AIE, $N_{\text{AIE}}$ is the total number of AIEs in the Computation Cores, and $\beta$ is the number of multiply-accumulate (MAC) operations that an AIE tile can perform each clock cycle. Using the ALU arrays to execute this task, the task is treated as SpDMM or SpMM, depending on which leads to lower execution cycle:
\begin{equation}
    t_{\text{ALU}} = \min \left(\alpha_{\text{min}}\frac{mnd}{pq},  \alpha_{\bm{X}_{i,:}} \alpha_{\bm{Y}_{:,j}} \frac{mnd}{p} \right)\times \frac{1}{\textit{f}_{\text{PL}}}
\end{equation}
where $\alpha_{\text{min}} = \min(\alpha_{\bm{X}_{i,:}}, \alpha_{\bm{Y}_{:,j}})$, and $\textit{f}_{\text{PL}}$ is the frequency of the ALU array.




\begin{table}[]
    \centering
    \renewcommand*{\arraystretch}{1.4}
    \caption{Performance Model}
        \label{Performance Model}
    
        \begin{tabular}{c |c| c c}
         \hline
          & $\mathcal{P}_{\text{AIE}}()$ &  \multicolumn{2}{c}{$\mathcal{P}_{\text{ALU}}()$} \\ 
         \hline\hline
         Primitives & GEMM & SpDMM & SpMM \\ 
         \hline
         MACs per Cycle & $N_{\text{AIE}} \times \beta$ & $pq$ & $p$ \\  [0.5ex]
         \hline
         Execution cycles & $\frac{mnd}{N_{\text{AIE}} \times \beta}$ & $\alpha_{\text{min}}\frac{mnd}{pq}$ & $\alpha_X\alpha_Y\frac{mnd}{p}$\\ 
         \hline
         Frequency & $\textit{f}_{\text{AIE}}$ & $\textit{f}_{\text{PL}}$ & $\textit{f}_{\text{PL}}$ \\  \hline
        \end{tabular}
        
\vspace{-4mm}
\end{table}
\begin{center}
\end{center}
\vspace{-3mm}

\vspace{-5mm}
\section{Experimental Results}
\label{sec:experimental-results}
\subsection{Implementation Details}

We implement the proposed accelerator on the AMD Xilinx Versal VCK5000 board.
We use Verilog HDL to develop ALU arrays on PL and Vitis to develop Task Scheduler on APU~\cite{xilinxcips} and AI Engine Compute Cores on AIEs. We integrate the complete system using Vitis. 
The Application Processing Unit (APU) - ARM Cortex-A72 runtime system uses C++ in AMD Xilinx Vitis Unified Software Platform (version 2021.2). 
For the AIE array, we implement 32 AIE CCs. Due to the limited on-chip memory, we implement 8 ALU arrays. In the performance model, each ALU array is configured with $p=8$ and $q=4$, determining the number of Multiply-Accumulate (MAC) operations per cycle.
The resource utilization of the overall system is summarized in Table \ref{tab:resource}. The ALU arrays, NoC, and AIE array operate at 297 MHz, 800 MHz, and 1 GHz, respectively.
We develop a cycle-accurate simulator for our architecture design and runtime system (Algorithm \ref{alg:Dynamic-Task-Scheduling}) to obtain the hardware execution time. In the cycle-accurate simulator, Ramulator \cite{kim2015ramulator} is used to simulate the performance of DDR memory. Also, we use the host processor to execute the preprocessing steps (see Section \ref{preprocessing}) and measure the preprocessing overhead.

\begin{table}[h]
\caption{Resource utilization of AIE and PL on VCK5000}
\vspace{-2mm}
\centering
    \renewcommand{\arraystretch}{1.3}

\label{tab:resource}

\begin{tabularx}{\linewidth}{|Z |YYY|}
\hline
\rowcolor[HTML]{C0C0C0}
& Tiles & \begin{tabular}[c]{@{}c@{}}AIE to NoC \\ Interface\end{tabular} & \begin{tabular}[c]{@{}c@{}}AIE to PL \\ Interface\end{tabular} \\ 
\rowcolor[HTML]{FFFFFF}
Overall AIE & 192 & 16 & 32 \\ 
\rowcolor[HTML]{EFEFEF}
Utilization & 48\% & 100\% & 100\% \\ \hline
\end{tabularx}

\vspace{1mm}

\begin{tabularx}{\linewidth}{|Z|YYYY|}
\hline
\rowcolor[HTML]{C0C0C0} 
& LUTS & DSPs & BRAMs & URAMs \\ 
\rowcolor[HTML]{FFFFFF} 
Overall PL & 776K & 1024 & 880 & 400 \\
\rowcolor[HTML]{EFEFEF} 
Utilization & 86.32\% & 52\% & 91\% & 86.4\% \\ \hline
\end{tabularx}
\vspace{-5mm}
\end{table}
%
%
%
%
%

\subsection{Experimental Setup}

\noindent \textbf{GNN Benchmarks}: 
We evaluate the performance of our design on four well-known GNN models: GCN \cite{kipf2016semi}, GraphSage \cite{hamilton2017inductive}, GIN \cite{xu2018powerful}, and SGC \cite{wu2019simplifying}. 



\vspace{0.1cm}
\noindent \textbf{Baselines}: We compare the performance of our accelerator against state-of-the-art CPU, GPU, and GNN accelerators, including HyGCN \cite{yan2020hygcn}, BoostGCN \cite{zhang2021boostgcn}, Dynasparse \cite{zhang2023dynasparse}, and H-GCN \cite{zhang2022h}. PyG and DGL are executed on Ryzen 3990x CPU and Nvidia RTX3090 GPU.  Details of the platforms are shown in Table \ref{platform}.

\begin{table}[h]
    \vspace{-2mm}
    \renewcommand{\arraystretch}{1.3}

\centering
\caption{Platform Specifications}
\vspace{-2mm}
\label{platform}
\resizebox{\columnwidth}{!}{%
\begin{tabular}{c|ccc}
    \toprule

    \rowcolor[HTML]{FFFFC7} Implementation & Platform & Frequency & \begin{tabular}[c]{@{}c@{}}DDR Memory\\ Bandwidth\end{tabular} \\ \midrule
    \rowcolor[HTML]{FFFFFF}CPU & Ryzen 3990x & 2.90 GHz & 107 GB/s \\
    \rowcolor[HTML]{EFEFEF} GPU & Nvidia RTX3090 & 1.7 GHz & 936.2 GB/s \\
    \rowcolor[HTML]{FFFFFF}HyGCN \cite{yan2020hygcn} & ASIC & 1 GHz & 256 GB/s \\
    \rowcolor[HTML]{EFEFEF} BoostGCN \cite{zhang2021boostgcn} & Stratix 10 GX & 250 MHz & 77 GB/s \\
    \rowcolor[HTML]{FFFFFF}Dynasparse \cite{zhang2023dynasparse} & Alveo U250 & 250 MHz & 77 GB/s \\
    \rowcolor[HTML]{EFEFEF} ACAP & VCK 5000 & \begin{tabular}{@{}c@{}}297 MHz (PL)\\ 1GHz (AIE)\end{tabular} & 102.4 GB/s \\ \bottomrule
\end{tabular}}
\end{table}

\noindent \textbf{Datasets}: We evaluate our design using several widely used datasets, including CiteSeer (CI) \cite{kipf2016semi}, Cora (CO) \cite{kipf2016semi}, PubMed (PU) \cite{kipf2016semi}, Flickr (FL) \cite{zeng2019graphsaint}, NELL (NE) \cite{yang2016revisiting}, and Reddit (RE) \cite{hamilton2017inductive}. We evaluate with 2-layer GNNs in \cite{kipf2016semi}, \cite{hamilton2017inductive}, \cite{xu2018powerful}, and \cite{wu2019simplifying}, where CI, CO, and PU have hidden layer dimensions of 16, while the hidden layer dimension of the remaining datasets is 128. Detailed dataset statistics are shown in Table \ref{dataset statistics}. 


\noindent \textbf{Performance Metrics}: We measure the \emph{hardware execution time}, which represents the duration from when the accelerator starts scheduling computations until it generates the final results. 
We also measure the \emph{preprocessing time}, which is the overhead of the data partitioning method (see Section \ref{subsec:system-overview}).

\begin{table}[h]
    \centering
    \caption {Dataset Statistics} 
    \label{dataset statistics}
    \vspace{-2mm}
    \renewcommand{\arraystretch}{1.3}
    \setlength{\tabcolsep}{0.65em}

    \resizebox{\columnwidth}{!}{%
    \begin{tabular}{c|cccccc}
    \toprule
    \rowcolor[HTML]{FFFFC7} 
    Dataset & Vertices & Edges & Features & Classes                     & \begin{tabular}[c]{@{}c@{}}Density of \\ $\bm{A}$\end{tabular} & \begin{tabular}[c]{@{}c@{}}Density of \\ input $\bm{H}$\end{tabular}  \\ \midrule
    \rowcolor[HTML]{FFFFFF}CO~\cite{kipf2016semi} & 2708 & 5429 & 2708 & 7 & 0.14\% & 1.27\%  \\
    \rowcolor[HTML]{FFFFFF}CI~\cite{kipf2016semi} & 3327 & 4732 & 3703 & 6 & 0.08\% & 0.85\%  \\
    \rowcolor[HTML]{FFFFFF}PU~\cite{kipf2016semi} & 19717 & 44338 & 500 & 3 & 0.02\% & 10\%  \\
    \rowcolor[HTML]{FFFFFF}FL~\cite{zeng2019graphsaint} & 89,250 & 899,756 & 500 & 7 & 0.01\% & 46\%\\
    \rowcolor[HTML]{FFFFFF}NE~\cite{yang2016revisiting} & 65,755 & 251,550 & 61278 & 186 & 0.0058\% & 0.01\%\\
    \rowcolor[HTML]{FFFFFF}Re~\cite{hamilton2017inductive} & 232,965  & 11$\times {10}^7$ & 602 & 41 & 0.21\% & 100\%                                                                                                                           \\ \bottomrule
    \end{tabular}
    }

\end{table}

\begin{figure}[h] 
\vspace{-5mm}

    \centering
        \includegraphics[width=\linewidth]{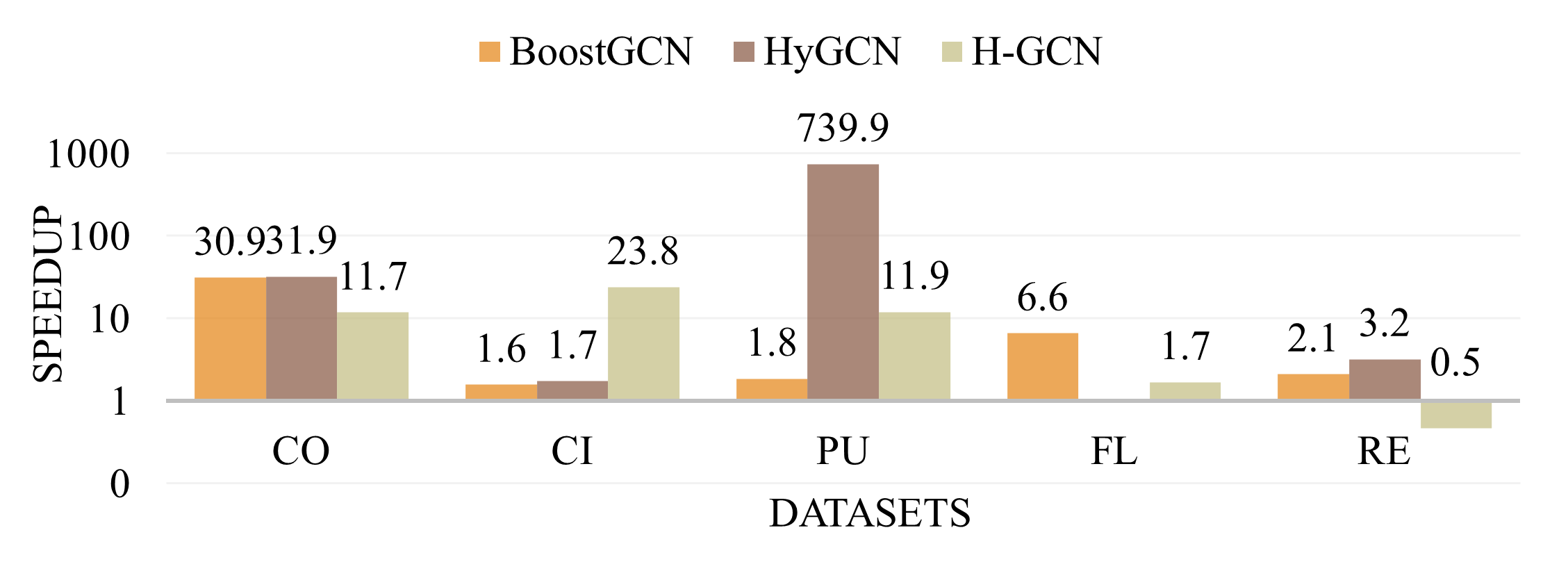} 
        \vspace{-6mm}
        \caption{Comparison of hardware execution time with state-of-the-art GNN accelerators}
        \label{speedup_over_gcnaccs}

\vspace{-2mm}
\end{figure}

\begin{table}[h]
\centering
\caption{FLOPs and data count exploiting sparsity in feature matrices (FMs) and adjacency matrix (AM) for GCN inference. "Sp. AM" refers to "Sparsity in AM only," and "Sp. AM + FMs" to "Sparsity in AM and FMs." The FLOPs Reduction Factor refers to the ratio of FLOPs count when exploiting Sp. AM + FM to the scenario of  Sp. AM only, while the Data Reduction Factor is the ratio for data count.}

\label{flops count}
\setlength{\tabcolsep}{0.7em}
\renewcommand{\arraystretch}{1.4}  

\resizebox{\columnwidth}{!}{%

\begin{tabular}{c|cccccc}
\hline

\rowcolor[HTML]{FE996B} 
                    & CO    & CI    & PU    & FL   & NE   & RE    \\ \hline
\rowcolor[HTML]{FFFFFF}
\#FLOPs Sp. AM     & 6.3E7 & 2.0E8 & 1.6E8  & 5.9E9 & 5.1E12 & 3.8E10  \\
\rowcolor[HTML]{EFEFEF} 
\#FLOPs Sp. AM + FMs & 1.2E6 & 2.1E6 & 1.8E7  & 2.8E9  & 5.3E10 & 3.7E10   \\ 
\rowcolor[HTML]{FFFFFF}
 FLOPs Reduction Factor& 48.6$\times$ & 95.5$\times$ & 8.8$\times$ & 2.1$\times$ & 9.7$\times$ & 1.0$\times$  \\ 
\hline
 \rowcolor[HTML]{EFEFEF} 
\#Data Sp. AM & 4.0E6 & 1.3E7 & 1.1E7 & 7.0E7  & 4.1E9  & 4.4E8 \\
\rowcolor[HTML]{FFFFFF} 
\#Data Sp. AM + FMs & 1.9E5 & 2.9E5 & 1.8E6 & 3.9E7 & 4.4E8 & 4.1E8 \\
 \rowcolor[HTML]{EFEFEF} 
 Data Reduction Factor & 20.9$\times$ & 43.5$\times$ & 6.0$\times$ & 1.8$\times$ & 9.2$\times$ & 1.1$\times$ \\
\hline

\end{tabular}%
}
\vspace{-3mm}
\end{table}

\begin{table*}[]   
    \centering
    \renewcommand{\arraystretch}{1.3}

    \caption {Comparison of hardware execution time with state-of-the-art CPU, GPU, FPGA, and ACAP implementations. The values are in $ms$ and rounded to the nearest hundredth (The best results are in \textbf{bold}, and the second best results are \underline{underlined}; OoM means out of GPU memory, and N/A means not available).}
    
    \label{inference time}

    \begin{tabularx}{\linewidth}{|c|c|YYYYYY|}
    \hline
    \multirow{2}{*}{Model} &\multirow{2}{*}{Platform}  &     \multicolumn{6}{c|}{Dataset}  \\
        \cline{3-8}
        & & CO       & CI       & PU       & FL       & NE       & RE   \\
        \hline\hline
        \multirow{9}{*}{GCN} & PyG-CPU & 2.10E+00 & 3.30E+00 & 8.70E+00 & 2.81E+02 & 1.54E+03 & 3.21E+04 \\
                            & PyG-GPU & 3.36E-01 & 3.76E-01 & 3.43E-01 & 7.02E+00 & 3.22E+01 & OoM      \\
                            & DGL-CPU & 1.90E+00 & 7.70E+00 & 7.20E+00 & 3.58E+01 & N/A      & 1.41E+02 \\
                            & DGL-GPU & 1.40E+00 & 1.40E+00 & 1.40E+00 & 2.10E+01 & N/A      & 5.07E+01 \\
                            & BoostGCN &2.90E-01 & 1.90E-02 & 1.60E-01 & 4.00E+01 & N/A   &1.90E+02 \\
                            & HyGCN & 3.00E-01 & 2.10E-02 & 6.40E+01 & N/A & N/A & 2.90E+02 \\
                            & H-GCN  & 1.10E-01 & 2.90E-01 & 1.03E+00 & 1.02E+01 & N/A     & {\textbf{4.18E+01}}\\
                            & Dynasparse &\textbf{4.70E-03 }& \textbf{7.70E-03} & \textbf{6.30E-02} & \underline{8.80E+00}& \textbf{2.90E+00} & 1.00E+02\\
                            & This paper &\underline{ 9.40E-03} &\underline{1.22E-02} &\underline{ 8.65E-02}& \textbf{6.10E+00} & \underline{5.20E+00}& \underline{9.10E+01} \\ \hline
        \multirow{6}{*}{GraphSage} & PyG-CPU & 1.36E+01 & 2.81E+01 & 4.15E+01 & 3.36E+02 & 2.13E+04 & OoM\\
                            & PyG-GPU & 7.30E-01 & 1.43E+00 & 1.69E+00 & 1.78E+01 & OoM & OoM     \\
                            & DGL-CPU & 3.42E+01 & 1.40E+01 & 2.43E+01 & 7.39E+01 & N/A & 3.39E+03 \\
                            & DGL-GPU & 8.61E-01 & 8.75E-01 & 8.37E-01 & 2.16E+01 & N/A & 4.45E+02\\
                            & Dynasparse & \underline{1.11E-01} & \underline{3.34E-01} & \underline{4.21E-01} & \underline{1.91E+01} & \underline{8.37E+02} & \underline{3.31E+02}\\

                            & This paper & \textbf{1.01E-01} & \textbf{2.51E-01} & \textbf{1.95E-01} & \textbf{1.91E+00} & \textbf{5.07E+02} & \textbf{2.81E+02} \\ \hline    
        \multirow{6}{*}{GIN} & PyG-CPU & 1.26E+01 & 3.27E+01 & 4.14E+01 & 5.05E+02 & 1.91E+04 & OoM\\
                            & PyG-GPU & 6.80E-01 & 1.46E+00 & 1.22E+01 & 1.73E+01 & OoM & OoM\\
                            & DGL-CPU & 6.00E+00 & 2.28E+01 & 1.82E+01 & 1.52E+02 & N/A & 3.39E+03\\
                            & DGL-GPU & 3.96E-01 & 4.30E-01 & 3.86E-01 & 1.95E+01 & N/A & 4.95E+02\\
                            & Dynasparse & \underline{1.08E-01} & \underline{3.29E-01} & \underline{3.71E-01} & \underline{1.21E+01} & \underline{8.37E+02} & \textbf{2.73E+02}\\

                            & This paper & \textbf{1.02E-01} & \textbf{2.52E-01} & \textbf{2.05E-01} & \textbf{7.61E+00} & \textbf{5.08E+02} & \underline{2.94E+02}\\ \hline        
        \multirow{6}{*}{SGC} & PyG-CPU & 2.44E+01 & 5.63E+01 & 7.63E+01 & 1.27E+03 & 4.32E+04 & OoM\\
                            & PyG-GPU & 1.08E+00 & 2.50E+00 & 3.01E+00 & 3.32E+01 & OoM & OoM\\
                            & DGL-CPU & N/A & N/A & N/A & N/A & N/A & N/A  \\
                            & DGL-GPU & N/A & N/A & N/A & N/A & N/A & N/A \\
                            & Dynasparse & \underline{2.67E+00} & \underline{8.70E-01} & \underline{2.34E+00} & \underline{1.27E+01} & \underline{8.84E+02} & \underline{5.05E+02}\\

                            & This paper & \textbf{1.22E-01} & \textbf{3.14E-01} & \textbf{3.18E-01} & \textbf{3.29E+00} & \textbf{7.82E+01} & \textbf{4.71E+02} \\ \hline

    \end{tabularx}
\end{table*}

\subsection{Comparison with State-of-the-art}

\noindent \textbf{Comparison with prior implementation on ACAP}:
We compare the performance of our implementation with a prior implementation on the same platform, H-GCN~\cite{zhang2022h}. Because we exploit data sparsity and utilize the heterogeneity of the platform and dynamically schedule the tasks to AIE and PL, we achieve an average of 9.9$\times$ speedup compared with H-GCN~\cite{zhang2022h}, as shown in Figure \ref{speedup_over_gcnaccs}.

This speedup is due to our exploitation of matrix sparsity in all the computation kernels, including feature aggregation and feature update. In Table \ref{flops count}, we provide a detailed analysis that shows a substantial reduction in both the number of floating-point operations (FLOPs) and the amount of data to be loaded, averaging 51$\times$ and 23.4$\times$, respectively, for the Planetoid datasets CO, CI, and PU. However, the reduction is comparatively smaller for FL and RE datasets (because the feature matrices of FL and RE have low sparsity. See Table \ref{dataset statistics}), resulting in a smaller speedup compared with H-GCN.

Additionally, while H-GCN demonstrates faster hardware execution time on the Reddit dataset, our proposed approach significantly reduces the preprocessing overhead, as discussed in Section \ref{preprocessing overhead}. Considering the end-to-end inference time, encompassing both the preprocessing overhead and the actual inference time, our method achieves a 6.6$\times$ speedup for the Reddit dataset.



\begin{figure*}[h] 
    \centering
        \includegraphics[width=\textwidth]{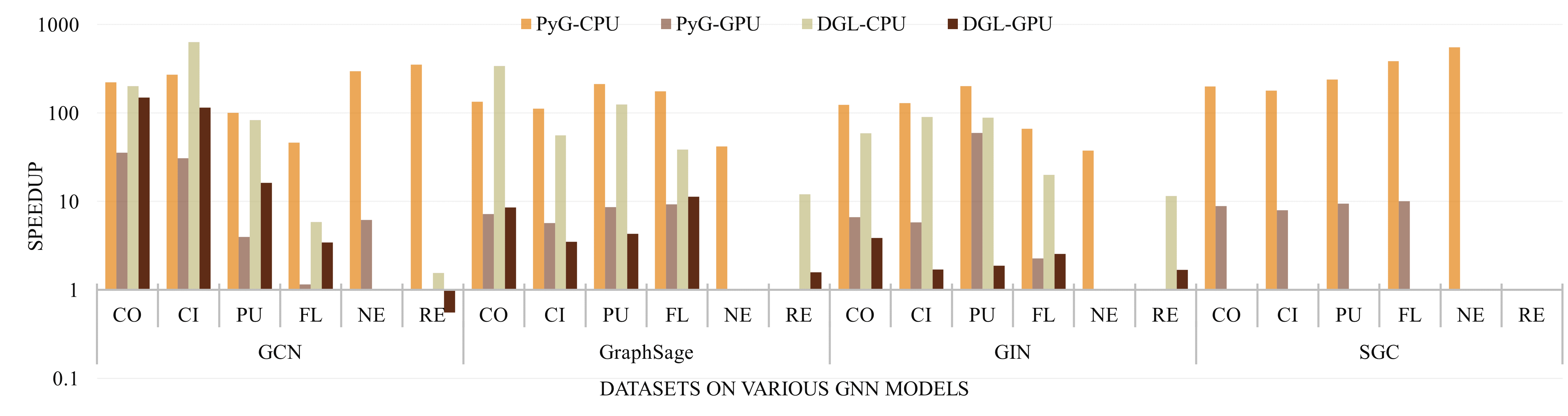} 
        \caption{Comparison of inference speedup over CPU and GPU (some speedups are not shown as those are OoM or N/A)}
        \label{speedup_over_cpu_gpu}
\end{figure*}

\noindent \textbf{Comparison with CPU and GPU}: We execute the same GNN models using state-of-the-art Pytorch Geometric (PyG)~\cite{fey2019fast} and Deep Graph Library (DGL)~\cite{wang2019deep} on a state-of-the-art CPU and GPU without exploiting data sparsity in feature matrix $\bm{H}$ and weight matrix  $\bm{W}$. The results are shown in Figure~\ref{speedup_over_cpu_gpu}; some results are not shown due to out of memory on the CPU/GPU. In summary, our implementation on ACAP achieves average speedup of 194.5$\times$, 12.9$\times$, 110.2$\times$, and 21.7$\times$ compared with PyG-CPU, PyG-GPU, DGL-CPU, and DGL-GPU, respectively. The achieved speedups are from exploiting the sparsity in GNN inference and the customized hardware architecture that can efficiently execute the sparse computation primitives (SpDMM, SpMM).

\vspace{0.1cm}
\noindent \textbf{Comparison with GNN Accelerators}:
The speedup compared with the state-of-the-art accelerators is shown in Figure \ref{speedup_over_gcnaccs}. The proposed design achieves an average speedup of 194.18$\times$ and 8.58$\times$  compared with HyGCN~\cite{yan2020hygcn} and BoostGCN~\cite{zhang2021boostgcn}. This is because our implementation utilizes the data sparsity in the vertex feature and input adjacency matrix, and AIE can efficiently execute the dense computation primitives (GEMM).
We also compare our design with Dynasparse \cite{zhang2023dynasparse}, which exploits data sparsity in GNN inference on FPGA. We achieve average speedup of 0.83$\times$, 2.90$\times$, 1.39$\times$, and 8.04$\times$ for GCN, GraphSage, GIN, and SGC models.  

The hardware execution time is summarized in Table \ref{inference time}, where the fastest hardware execution time for various models and datasets  is highlighted in \textbf{bold}, and the second fastest time is \underline{underlined}. Our design achieves the best or second-best hardware execution time across all models and datasets.

\begin{table}[h]
    \centering
    \caption {Hardware execution time on ACAP ($ms$) using PL only and using PL + AIE} 
    \label{ACAP time comparison}
    \renewcommand{\arraystretch}{1.2}

    \scalebox{1}{
    \setlength{\tabcolsep}{0.3em}

        \begin{tabular}{|c | c c c c c c |} 
         \hline
         {Dataset} & {CO} & {CI} & {PU} & {FL} & {NE} & {RE} \\  
         \hline\hline
            PL Only &	2.45E-1&	7.26E-1&	6.55E-1&	2.09E+1&	5.02E+2&	3.52E+2\\[0.5ex]
         \hline
            PL + AIE &	9.40E-3&	1.22E-2&	8.65E-2&	6.10E+0&	5.20E+0&	9.10E+1\\[0.5ex]
         \hline
         
    \end{tabular}
}
\vspace{-1mm}
\end{table}

\subsection{Exploring the heterogeneity of ACAP}
\label{heter_Acap}
Table \ref{ACAP time comparison} compares the hardware execution time of the proposed accelerator design (PL + AIE) and the PL accelerator design (PL Only) for various datasets for GCN inference. The results highlight the significant speedup achieved by leveraging the heterogeneity of the ACAP device. On the average, the PL + AIE design achieves a speedup of 32.9$\times$ compared with the PL-only design. 
The improvement is due to the architecture of the AIE array that provides high parallelism when processing GEMM primitives, while the PL can efficiently compute the sparse primitives (SpDMM, SpMM).

\begin{table}[h!]
    \centering
    \caption {Hardware execution time ($ms$) using various numbers of AIE tiles assuming sufficient external memory bandwidth. (192 and 384 are the number of AIE tiles used.)} 
    \label{scaled up version}
    \renewcommand{\arraystretch}{1.2}

    \scalebox{1}{
    \setlength{\tabcolsep}{0.3em}

        \begin{tabular}{|c | c c c c c c |} 
         \hline
         {Dataset} & {CO} & {CI} & {PU} & {FL} & {NE} & {RE} \\  
         \hline\hline
            Current(192) &	9.40E-3&	1.22E-2&	8.65E-2&	6.10E+0&	5.20E+0&	9.10E+1\\[0.5ex] 
         \hline
            Scaled(384) &	9.40E-3&	1.22E-2&	8.65E-2&	2.53E+0&	4.25E+0&	7.97E+1\\[0.5ex] 
         \hline
         
    \end{tabular}
}
\vspace{-5mm}
\end{table}


The board has limited external memory access bandwidth, so our current design uses only 32 AIE CCs (192 tiles). Increasing the number of AIE CCs will not proportionally increase the peak performance (\# of AIE CCs * \#MACs/cycle) as the computation would become memory bound. However, we simulate a scenario with double the AIE CCs, using 384 of 400 AIE tiles, assuming sufficient external memory access bandwidth to support all the AIE CCs.
Table \ref{scaled up version} shows that for the larger datasets (FL, NE, RE), increasing the number  of tiles shows a speedup in hardware execution time. However, our hardware execution time on RE is still slower than H-GCN for RE as SpDMM dominates the overall performance on RE. While H-GCN utilizes AIEs to execute SpDMM, our approach uses PL only.  Despite each ALU array being more efficient than one AIE CC at computing sparse primitives, the AIE has superior overall peak performance on SpDMM than PL-based design. Therefore, our hardware execution time is slower than H-GCN on RE dataset.

\subsection{Analysis of Preprocessing and Runtime System Overhead} 
\begin{figure}[h] 
    \centering
        \includegraphics[width=\linewidth]{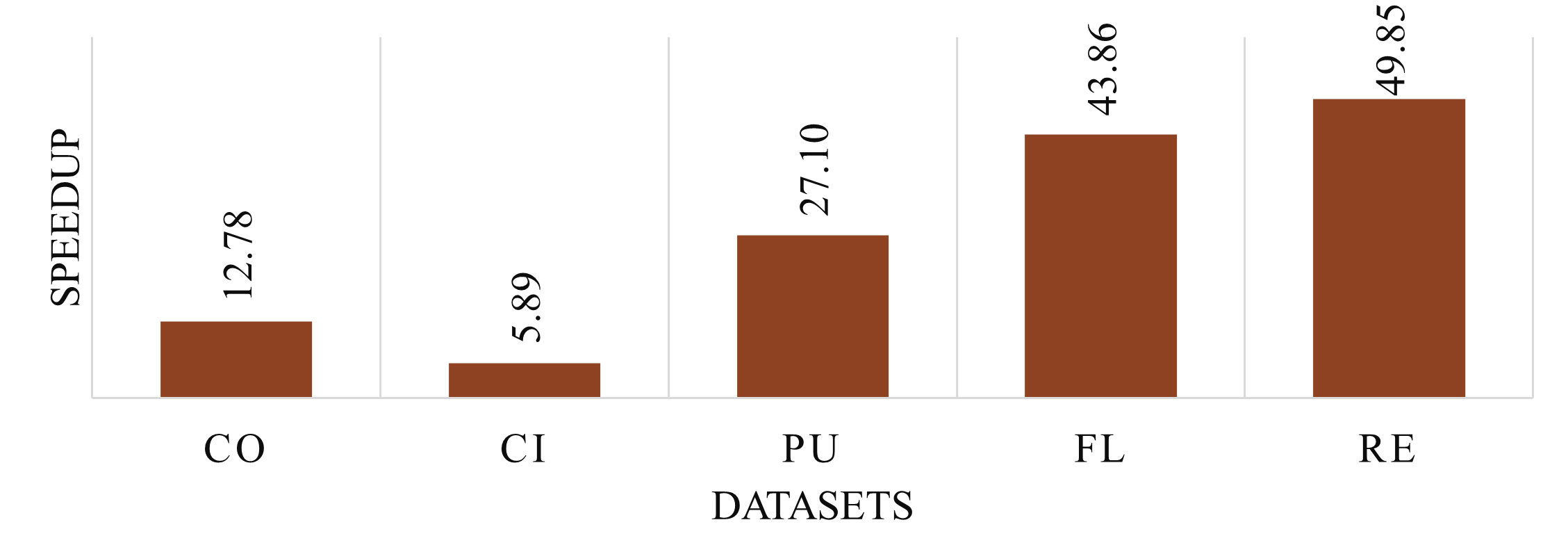} 
        \vspace{-3mm}
        \caption{Comparison of preprocessing time with H-GCN~\cite{zhang2022h}}
        \label{speedup in preprocessing}
        \vspace{-3mm}
\end{figure}

\noindent \textbf{Preprocessing Overhead}: 
\label{preprocessing overhead}
We evaluate the overhead of preprocessing, detailed in Section \ref{preprocessing}. This involves data partitioning on the host processor (Intel Xeon Gold CPU with 32 cores at 2.9 GHz) only once before the inference tasks start. The overhead of partitioning was smaller than the preprocessing time of the state-of-the-art GCN Accelerator on ACAP, H-GCN (which used Intel Xeon Gold with 56 CPU cores \cite{zhang2022h}). Figure \ref{speedup in preprocessing} shows our speedups in preprocessing time.


\vspace{-3mm}

\vspace{0.1cm}
\noindent \textbf{Runtime System Overhead}: The runtime system overhead corresponds to the execution time of Algorithm \ref{alg:Dynamic-Task-Scheduling}, performed on the Arm Cortex-A72 APU running at 1.7 GHz. After the initial tasks assignment, the runtime system overhead can be overlapped by concurrently analyzing and scheduling the tasks on AIE CCs and ALU arrays while they are working on the previously assigned tasks. The time it takes for the runtime system to analyze and schedule the initial tasks is less than 1\% of the total hardware execution time.


%

\section{Conclusion and Future Work}
In this paper, we proposed a hardware accelerator that utilized the heterogeneity of Versal Architecture to exploit the data sparsity to accelerate GNN inference. The proposed system that dynamically maps tasks on PL and AIE leads to the speedup of 3.9-96.7× compared to PL-only implementation for GCN inference. The proposed design achieves 162.42×, 17.01×, 9.90×, and 27.23× average speedup compared with the state-of-the-art implementations on CPU, GPU, other ACAP, and other GNN accelerators, respectively. 


Currently, the limited PL resources become a bottleneck. This restricts the number of ALU arrays that can be compiled, causing sparse primitives to dominate the overall execution time for some datasets. In the future, we plan to implement more resource-efficient ALU arrays and expand the use of AIE for sparse computations such as SpDMM and SpMM. This strategy would allow the AIE array to support sparse computations when the ALU arrays are fully utilized.

\section*{Acknowledgment}
This work is supported by the National Science Foundation under grants CCF-1919289 and OAC-2209563. Equipment and support by AMD AECG are greatly appreciated.









\bibliographystyle{IEEEtran}
\newpage
\bibliography{reference}

\end{document}